\documentclass{svproc}
\pdfoutput=1
\usepackage[utf8]{inputenc}
\usepackage{color,graphicx,hyperref,cite,multirow,dsfont,slashed,multirow}
\hypersetup{bookmarks=true,unicode=true,pdftoolbar=true,pdfmenubar=true,
 pdffitwindow=false,pdfstartview={FitH}, pdftitle={Zprime},
 pdfauthor={J.~Araz}, pdfsubject={Zprime},
 pdfcreator={J.~Araz}, pdfproducer={},
 pdfkeywords={Zprime},
 pdfnewwindow=true,colorlinks=true,
 linkcolor=blue,citecolor=magenta,filecolor=magenta,urlcolor=cyan}
\usepackage{url}

\newcommand{\be}{\begin{equation}}
\newcommand{\ee}{\end{equation}}
\def\bsp#1\esp{\begin{split}#1\end{split}}

\newcommand\bpm{\begin{pmatrix}}
 \newcommand\epm{\end{pmatrix}}

\renewcommand{\figureautorefname}{Fig.}

\def\sectionautorefname~#1\null{Sec.~#1\null}
\def\subsectionautorefname~#1\null{subsec.~#1\null}
\def\figureautorefname~#1\null{Fig.~#1\null}
\def\tableautorefname~#1\null{Table~#1\null}
\def\equationautorefname~#1\null{Eq.~(#1)\null}

\begin{document}
\mainmatter              
\title{Leptophobic $ Z^\prime $ in supersymmetry and where to find them}
\titlerunning{Leptophobic $ Z^\prime $ in supersymmetry}  
\author{Jack Y. Araz}
\authorrunning{J.Y. Araz} 
\institute{Concordia University 7141 Sherbrooke St. West, Montr\'{e}al, QC, Canada H4B 1R6\\
 \email{jack.araz@concordia.ca}}

\maketitle              

\begin{abstract}
This report presents possible loopholes in heavy $ Z^\prime $ searches which are concentrated around high mass resonances. We present a scenario where massive neutral vector boson $ Z^\prime $, predicted by $ E_6 $-inspired $ U(1)^\prime $ models, can become leptophobic due to kinetic mixing between $ Z-Z^\prime $. Such a scenario can avoid high-mass dilepton resonance searches. Furthermore, we propose a channel through supersymmetric cascade decays to observe leptophobic $Z^\prime$. We show that it is possible to reach up to $ 8\sigma $ at $ \sqrt{s} = 14 $~TeV and 3~ab$ ^{-1} $ integrated luminosity. This report follows the steps of the work that has been conducted in ref.~\cite{Araz:2017wbp}.
\end{abstract}
%
\section{Introduction}
Heavy neutral gauge boson, $ Z^\prime $, can be obtained by extending the Standard Model~(SM) gauge structure with a $ U(1) $ gauge group. Such models have been widely studied under Grand Unified scheme~\cite{RevModPhys.81.1199}. Similarly, under supersymmetric realisations, $Z^\prime$ can emerge from $U(1)$ extended minimal supersymmetric extension of the SM~(UMSSM). It can also be achieved in so-called Sequential Standard Model~(SSM) where $ Z^\prime $, and $ W^\prime $, bosons have same couplings to fermions as their SM counterparts. 

Search for heavy neutral vector bosons in LHC has been extensively studied by experimental collaborations due to its relatively familiar resonance signature~\cite{Aaboud:2017buh, Sirunyan:2016iap, CMS:2016abv, PhysRevD.96.052004, CMS:2019oju, Aad:2019fac}. These searches set the mass bound for the $ Z^\prime $ assuming that it can only decay through the SM particles. These studies focus on high-mass dilepton or dijet resonances. In case of the dijet resonance studies at $\sqrt{s} = 13$~TeV,  ATLAS collaboration~\cite{PhysRevD.96.052004} presented the $Z^\prime$ mass bounds around $2.1-2.9$~TeV for 37 fb$^{-1}$ integrated luminosity and CMS collaboration~\cite{Sirunyan:2016iap} around 2.7 TeV for 12.9~fb$^{-1}$. In case of dilepton resonance studies, ATLAS collaboration presented the mass bounds at $3.8-4.1$~TeV for 36.1~fb$^{-1}$~\cite{Aaboud:2017buh} and CMS collaboration presented at 3.5~TeV for 13~fb$^{-1}$ integrated luminosity\footnote{It is important to note that these limits are recently updated. See ref.~\cite{CMS:2019oju} for dijet resonances at 137 fb$^{-1}  $ and ref.~\cite{Aad:2019fac} for dilepton resonances at 139 fb$ ^{-1} $.}. We investigate possible loopholes within the specific scenarios to lower these mass limits and also propose a leptophobic framework for $ Z^\prime $ to search for a possible signal through supersymmetric decay channels in HL-~\&~HE-LHC.

The rest of this report is organised as follows; in \autoref{sec:UMSSM}, we introduced the theoretical properties of UMSSM and discussed current mass bounds on $ Z^\prime $ at the time when this study conducted. In the following \autoref{sec:HLLHC}, we discussed possible supersymmetric signatures to observe a leptophobic $ Z^\prime $ in HL-~\& HE-LHC and finally we conclude in \autoref{sec:conclusion}.

\section{Theoretical framework and $ Z^\prime $ mass bounds}\label{sec:UMSSM}

Within a possible minimal representation in supersymmetry (SUSY), $ Z^\prime $ bosons can arise from the breaking of rank-6, $ E_6 $ group, which is greatly motivated by Grand Unified Theories. Among the various possible breaking schemes of $ E_6 $ exceptional group, so-called UMSSM framework being the $ U(1)^\prime $ extended MSSM, arises from the breaking through $ SO(10) $. The two-step breaking structure creates separate $ U(1) $ gauge groups \cite{SLANSKY19811,Benavides:2018fzm,Araz:2017qcs}
\begin{eqnarray}
E_6 \to SO(10)\otimes U(1)_\psi \to (SU(5)\otimes U(1)_\chi)\otimes U(1)_\psi,\nonumber
\end{eqnarray}
where these $U(1)$'s can mix via $ \theta_{E_6} $ angle and generate $ U(1)^\prime $,
\begin{eqnarray}
U(1)^\prime = U(1)_\psi \cos\theta_{E_6} - U(1)_\chi\sin\theta_{E_6}\ .\nonumber
\end{eqnarray}
Here $ SU(5) $ group further breaks into MSSM gauge structure. Alongside with such vector boson, extending MSSM framework with a $ U(1) $ gauge provides an anomaly free gauge structure. This can prevent the R-parity violating terms from the Lagrangian and eliminates short-lived proton possibility which occurs due to $ SU(5) $ breaking. 

In $ E_6 $, the matter sector of UMSSM arises from the breaking of $\mathbf{27} $-plet vector representation in to $ \mathbf{16}\oplus\overline{\mathbf{10}}\oplus1 $ where $\mathbf{16}  $ further decomposes to the matter fields, $ \overline{\mathbf{10}} $ to electroweak Higgs fields and exotic quarks and singlet, $1$, becomes $U(1)^\prime$ singlet superfield, $ S $. In addition to the MSSM's large particle content, UMSSM accommodates extended slepton, sneutrino and squark sector up to 6 supersymmetric particles in each family. Also, breaking $ SO(10) $'s $ \textbf{16} $-plet vector representation provides a right-handed neutrino superfield which generates right-handed neutrino and its supersymmetric partner in this framework. The Higgs sector is consists of two charged $ H^{\pm} $ and four neutral scalar bosons, which are a pseudoscalar $ A $ and three neutral scalars $ h $, $ H $ and $ H^\prime $ where $ h $ and $ H $ are being MSSM-like Higgs. This spectrum also creates an extended gaugino sector with a total of six neutralinos where chargino sector remains unchanged.

At tree level, the squared mass of $ Z^\prime $ boson is given by
\begin{eqnarray}
M^2_{Z^\prime} = g^\prime ({Q^\prime_S}^2 v^2_S + {Q^\prime_{H_u}}^2 v^2_{H_u} +{Q^\prime_{H_d}}^2 v^2_{H_d})\ , \nonumber
\end{eqnarray}
where $ g^\prime $ being the coupling constant of the $ U(1)^\prime $ gauge group, $ Q^\prime_i $ is corresponding $ U(1)^\prime $ charge of singlet $ s $, doublet $ H_u $ and doublet $ H_d $ superfields and $ v_i $ being their respective vacuum expectation values. The existence of the $ U(1)^\prime $ group leads to extra D-~\& F-terms which contributes to the masses of the supersymmetric partners and Higgses. Large bounds of $ Z^\prime $ boson requires large $ v_S $ which further leads to a heavy particle spectrum especially in the Higgs sector. Thus, requiring massive $ Z^\prime $ also causes massive particle spectrum, which can be tuned via corresponding softmasses.

\begin{figure}
 \centering
 \includegraphics[scale=0.29]{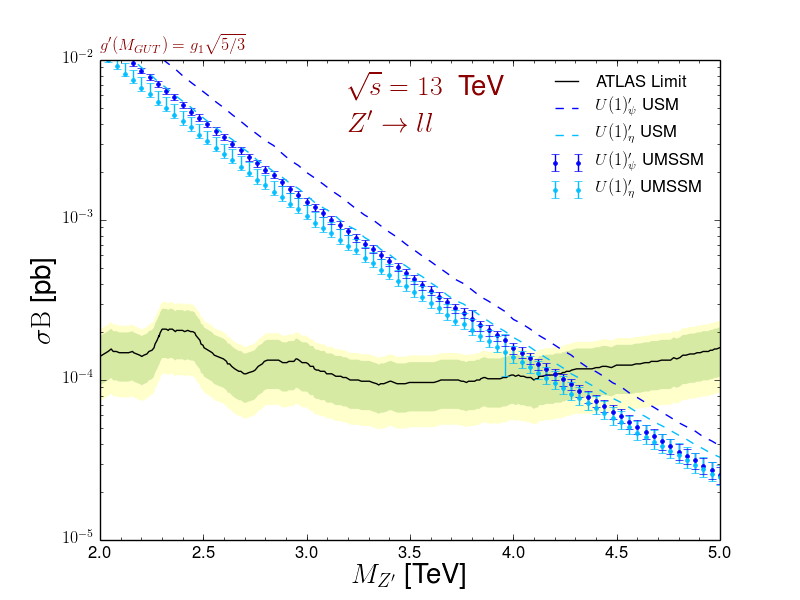}
 \includegraphics[scale=0.29]{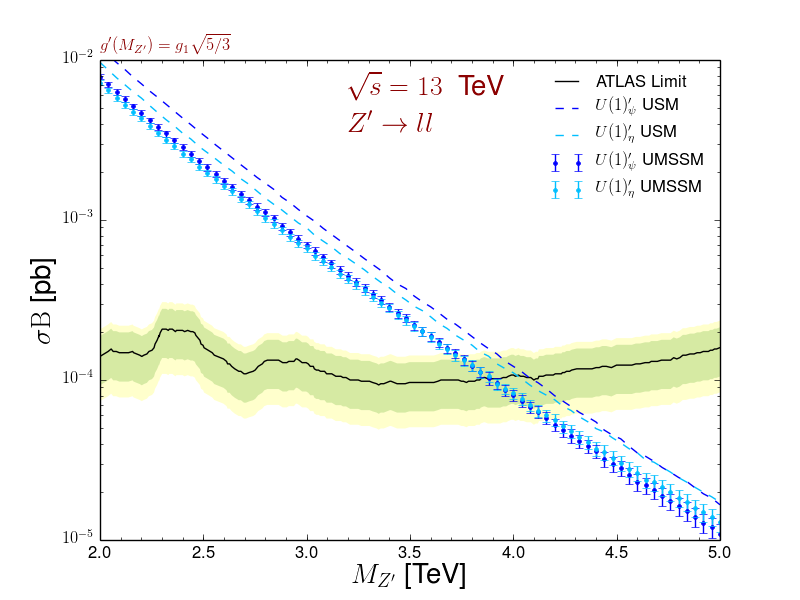}
 \caption{\it The effect of symmetry breaking scale in both inclusive (UMSSM) and noninclusive (USM) supersymmetric particle cases are presented. The left panel shows $ Z^\prime $ mass versus its corresponding cross section times branching ratio to dilepton final states where SUSY breaking occurring at GUT scale. The dotted line represents UMSSM and dashed line USM realisations, and the error bars shows the variation caused by theoretical uncertainties. The right panel shows the same for SUSY breaking occurring at $ Z^\prime $ mass scale. These results are presented in ref.~\cite{Araz:2017wbp}.\label{fig:gp_scale}}
\end{figure}

The interaction Lagrangian of the fields $ \hat{B}_\mu $ and $ \hat{Z}^\prime_\mu $ with the fermions are suppressed by their corresponding coupling and charge, $ g_1Y_i $ and $ g^\prime Q^\prime_i $ respectively where $ i $ represents the corresponding fermionic field in the interaction, $ g_1 $ and $ g^\prime $ are $ U(1)_Y $ and $ U(1)^\prime $ couplings respectively. The typical $ E_6 $-inspired relation between $ U(1)_Y $ and $ U(1)^\prime$ couplings are taken to be $ g^\prime/g_1 = \sqrt{5/3}$.  In ref. \cite{Araz:2017wbp}, it has been shown that shifting this relation from GUT scale to $ Z^\prime $-mass scale, and abandoning grand unification assumption, and adding the possibility to decay through supersymmetric particles, can reduce $ Z^\prime $ mass limit up to 300 GeV depending on $ \theta_{E_6} $ angle~\cite{Corcella:2012dw, Corcella:2014lha}. Breaking SUSY sector in a higher scale generates enough room for $ U(1)^\prime $ coupling to grow via renormalisation group equations which leads to larger $g^\prime$ and higher production amplitudes for $ Z^\prime $. Thus, breaking SUSY in a lower energy scale limits the growth in $g^\prime$ and reduces the production cross-section, which leads to relatively less stringent mass bounds for $Z^\prime$.  In \autoref{fig:gp_scale} we compared high-mass dilepton limit from ATLAS collaboration \cite{Aaboud:2017buh} to our results, sampling $ \theta_{E_6} $ at $ -0.79\pi $ ($ U(1)^\prime_\eta $) and $ 0 $ ($ U(1)^\prime_\psi $), where $ Z^\prime $ production has been calculated in next-to-leading order (NLO) precision. In both panels, the y-axis shows the production cross section with its branching fraction to dilepton final states where $ Z^\prime $ mass varied between $ 2 - 5 $ TeV within the bounds of narrow width approximation. Panels show the effect of different breaking scales where for the left panel $ g^\prime/g_1 $ relation is taken at GUT scale, and in the right panel, it downgraded to $ Z^\prime $ mass scale. 
The dashed line shows the case where $ Z^\prime $ is only allowed to decay into the SM particles (USM), and the dotted line shows the case with SUSY particles added (UMSSM). The error bars represent the theoretical uncertainties, scale and parton distribution function (PDF) uncertainties added in quadrature. The calculation has been performed via {\sc MadGraph}~\cite{Alwall:2014hca}. The left panel of \autoref{fig:gp_scale} reveals that, when the relation between couplings are assumed to be taken at GUT scale, the addition of supersymmetric partners to the possible decay channels improves the mass limits around 20\% and reduces it up to 4 TeV. Moving this relation to a lower scale in the right panel reduces the mass limits up to 3.8 TeV, where still preserving the difference between USM and UMSSM.

Although reducing production amplitudes and adding supersymmetric particles as possible decay channels create a considerable deviation from the current bounds, it is also possible to propose an altogether leptophobic $ Z^\prime $ to avoid the bounds coming from Drell-Yan modes. Leptophobic $Z^\prime$ can be achieved by mixing $ U(1)_Y $ and $ U(1)^\prime $ gauge fields. The mixing can be delivered in two folds. First, there can be mixing between $ Z $ and $ Z^\prime $ mass eigenstates. This, however, is constrained by electroweak precision tests to be at the $ \mathcal{O}(10^{-3}) $ \cite{Erler:2002pr}. Secondly, a kinematical mixing between $ Z $ and $ Z^\prime $ can be introduced through an angle $ \chi $,
\begin{eqnarray}
\mathcal{L}_{\rm kin} \subset -\frac{\sin\chi}{2} \hat{B}^{\mu\nu}\hat{Z}^\prime_{\mu\nu}\ , \nonumber
\end{eqnarray}
where $ \hat{B}^{\mu\nu} $ and $ \hat{Z}^\prime_{\mu\nu} $ are the $ U(1)_Y $ and $ U(1)^\prime $ boson field strength tensors respectively. This mixing further modifies the interaction vertices between $ Z^\prime $ and other fermions. The interaction term can be written in terms of canonical (diagonal) kinetic terms, $ Z^\prime_\mu $, as,
\begin{eqnarray}
\mathcal{L}_{\rm int} \subset -g^\prime \bar{\psi}_i\gamma^\mu\ \left(\frac{Q^\prime_i}{\cos\chi} - \frac{g_1}{g^\prime}Y_i \tan\chi\right)\ Z^\prime_\mu\psi_i\ ,\nonumber
\end{eqnarray}
where $ \psi_i $ is corresponding fermionic field and $ Y_i $ and $ Q^\prime_i $ being their $ U(1)_Y $ and $ U(1)^\prime $ charges. For simplicity, we limited the Lagrangian with the $ Z^\prime $ portion. The term in the paranthesis can be represented as modified $ U(1)^\prime $ charge, $ \bar{Q}_i $. To end up with a leptophobic $ Z^\prime $, $ \bar{Q}_i $ needs to be zero for left and right lepton superfields \cite{Babu:1996vt}. As shown in \cite{Araz:2017wbp}, left and right leptonic superfield charges can be set to zero, $ \bar{Q}_L=\bar{Q}_E=0 $, only around 
\begin{eqnarray}
\theta_{E_6} = (n-0.79)\pi\quad{\rm for}\quad n = 0,1,2\dots\quad, \label{eq:thE6_leptophobic}
\end{eqnarray}
which requires $ \sin\chi \cong (-1)^{n+1}0.3 $, assuming $ g^\prime/g_1 = \sqrt{5/3}$ ratio is satisfied at $ Z^\prime $ mass scale.

\begin{figure}[!h]
 \centering
 \includegraphics[scale=0.29]{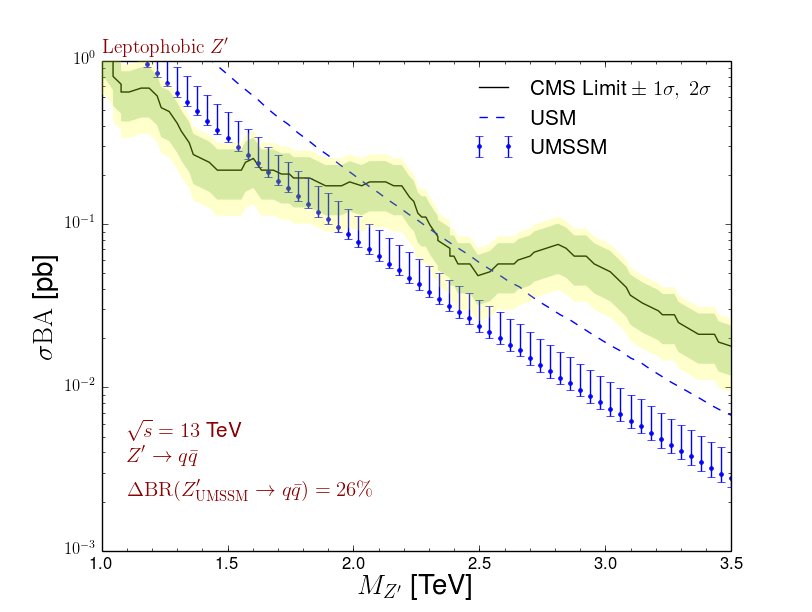}
    \caption{\it Plot shows leptophobic $ Z^\prime $ mass versus production cross-section and dijet branching ratio with corresponding acceptance rate. The dashed line shows the USM sample, which assumes that $ Z^\prime $ can only decay through the SM particles, and the dotted line shows UMSSM, including all supersymmetric particles. Error bars include scale and PDF variations as well as the variation in the dijet decay rate per sample. The CMS limit is taken from ref. \cite{Sirunyan:2016iap}. \label{fig:dijet}}
\end{figure}

Such leptophobic $ Z^\prime $ can only be bounded via high-mass dijet searches. It has been shown, in ref. \cite{Sirunyan:2016iap}, that these bounds are much weaker compared to Drell-Yan signature. In \autoref{fig:dijet} we performed a similar comparison as above where this time the leptophobic $ Z^\prime $ mass varied to compare with CMS collaboration's dijet limits \cite{Sirunyan:2016iap}. In order to zoom into the respectively low mass region, we varied $ Z^\prime $ mass between $ 1-3.5 $ TeV and plotted against the multiplication of its production cross section, branching ratio to dijet final states and the acceptance rate stated in ref. \cite{Sirunyan:2016iap}. Again, the dashed line assumes that $ Z^\prime $ can only decay to the SM particles (USM) and dotted line includes its possible decay channels to SUSY particles as well (UMSSM). In addition to the theoretical uncertainties, the error bars also shows the variation in the total decay rate to dijet final states. Such construction reduces the $ Z^\prime $ mass limits to 1.6 TeV and creates a considerably large deviation between USM and UMSSM depending on the dijet branching fraction.

\section{Leptophobic $ Z^\prime $ in HL-~\&~HE-LHC}\label{sec:HLLHC}

\begin{figure}[!h]
 \centering
 \includegraphics[scale=0.4]{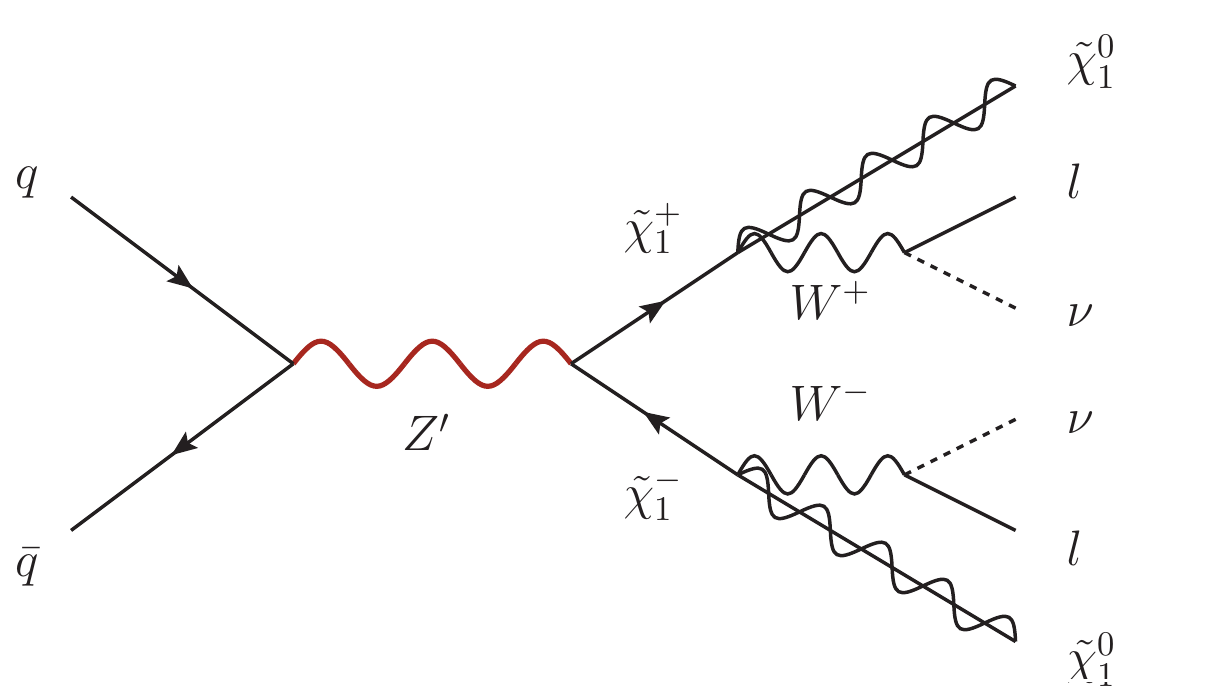}
 \caption{\it Choosen analysis signal where $ Z^\prime $ production and its decay through two lightest charginos. The figure has been produced with the help of the \textsc{JaxoDraw} package~\cite{Binosi:2008ig}.\label{fig:chaprod}}
\end{figure}

A leptophobic $ Z^\prime $ can still be observed through leptonic final states, but instead of direct decays, leptons can be produced through a cascade decay of supersymmetric particles. In \autoref{fig:chaprod}, we proposed a possible scenario where two charginos are pair produced via on-shell $ Z^\prime $ and they further decay into missing energy and dileptons through $ W $.  Such scenario might generate its background through heavy Higgs production and their decay to charginos. However, we observed that the branching fraction of such decay is negligible compared to heavy $ Z $ boson's decay rate. Also, neutralino pair production or chargino-neutralino associative production through $ Z^\prime $ has been observed to have small branching fraction compared to chargino pair production.
To test this hypothesis, we choose two UMSSM benchmarks to study possible observability of such leptophobic scenarios. Following ref.~\cite{Araz:2017wbp}, we chose two benchmarks using \autoref{eq:thE6_leptophobic}, where for \textbf{BM~I} $ n $ chosen to be $0$ and for \textbf{BM~II} $ n $ chosen to be $ 1$.  A conservative, $2.5$ TeV $ Z^\prime $ mass has been chosen for both benchmarks where other parameters are shown in \autoref{tab:parabm}. The particle mass spectrum has been calculated in \textsc{SARAH}~\cite{Staub:2013tta} interfaced with \textsc{SPheno}~\cite{Porod:2011nf,Porod:2003um}. The particle masses and decay tables can be found in ref.~\cite{Araz:2017wbp} which are not quoted here. The $ Z^\prime $ branching ratio to chargino pairs are about 2\% (6\%) for \textbf{BM~I} (\textbf{BM~II}). To ensure 100\% branching ratio of chargino to LSP, the mass splitting between two particles are chosen to be slightly larger than $ W $ mass.
\begin{table}
 \renewcommand{\arraystretch}{1.2}
 \begin{center}
  \begin{tabular}{c||c|c|c|c|c}
   Parameter &
    $\tan\beta$ & $\mu_{\rm eff}$ [GeV]  & $M_0$ [TeV] &  $M_1$ [GeV] & $ M_{\tilde{\chi}^\pm_1} $ [GeV]\\ \hline \hline
   {\bf BM I} & 9.11 & 218.9  & 2.6 & 106.5 & 344\\ \hline
   {\bf BM II} & 16.08 & 345.3 & 1.9 & 186.7 & 178\\[.1cm]
   \multicolumn{6}{c}{}\\
   Parameter &
   $M_2$ [GeV] & $M_3$ [TeV] & $M_1^\prime$ [GeV] & $A_0$ [TeV] & 
   $A_\lambda$ [TeV] \\ \hline \hline
   {\bf BM I} &
   230.0 & 3.6 & 198.9 & 2 & 5.9  \\ \hline
   {\bf BM II} &
   545.5 & 5.5 & 551.7 & 1.5 & 5.1 
  \end{tabular}
  \caption{\label{tab:parabm}
   UMSSM parameters for the reference points
   {\bf BM I} and {\bf BM II}.}\end{center}
\end{table}

 The amplitudes of the process are computed at NLO precision for a centre-of-mass energy $ \sqrt{s} = 14 $ TeV, using \textsc{MadGraph5\_amc@NLO}~\cite{Alwall:2014hca}  with the NLO set of NNPDF 2.3 parton densities. The production cross section of $ Z^\prime $ calculated as 120 fb. Prepared matrix-level events are hadronized and showered in \textsc{Pythia}~8~\cite{Sjostrand:2014zea} and the detector response has been simulated in \textsc{Delphes}~\cite{deFavereau:2013fsa} using the \textsc{Snowmass} parametrization~\cite{Berggren:2013hda,Anderson:2013kxz}. The jets are clustered using \textsc{FastJet} program~\cite{Cacciari:2011ma} with \texttt{antikT} algorithm using $R=0.6$.
 
 For the preselection, jets (leptons) are required to have minimum 40 (20) GeV transverse momentum and $ |\eta| < 2.4\ (1.5)$. Leptons within $ \Delta R<0.4 $ to a hard jet are omitted. Only two muons are allowed in the final state with 15\% hadronic activity within a cone radius $ R=4 $ and the angular separation in the transverse plane between two leptons are required to be greater than $2.5$. To isolate the dilepton final state, all jets are vetoed. The leading and second-leading leptons are required to have at least 300 GeV and 200 GeV transverse momentum respectively. Finally, the missing transverse energy has been required to be greater than 100 GeV. We calculated the significance of the signal over the background using two different approaches. One being standard $s = S/\sqrt{B+\sigma^2_B} $ and other begin Asimov significance~\cite{Cowan:2010js} given as;
\begin{eqnarray}
Z_A=\sqrt{ 2\left(
 (S+B)\ln\left[\frac{(S+B)(S+\sigma^2_B)}{B^2+(S+B)\sigma^2_B}\right] -
 \frac{B^2}{\sigma^2_B}\ln\left[1+\frac{\sigma^2_BS}{B(B+\sigma^2_B)}\right]
 \right)}\ ,\nonumber
\end{eqnarray}
where $ S $ being the number of signal events, $ B $ number of the background events and $ \sigma_B $ is the uncertainty over the background. For 10\% systematic uncertainty on the background, we calculated $ s = 4.2\sigma $, $ Z_A = (3.5\pm 0.9)\sigma$ for \textbf{BM~I} and $ s = 8.14\sigma $, $ Z_A = (6.06\pm 1.18)\sigma$ for \textbf{BM~II} at 3 ab$ ^{-1} $. The details of the analysis alongside with the differential distributions can be found in ref.~\cite{Araz:2017wbp}.

\begin{figure}[!h]
 \centering
 \includegraphics[scale=0.29]{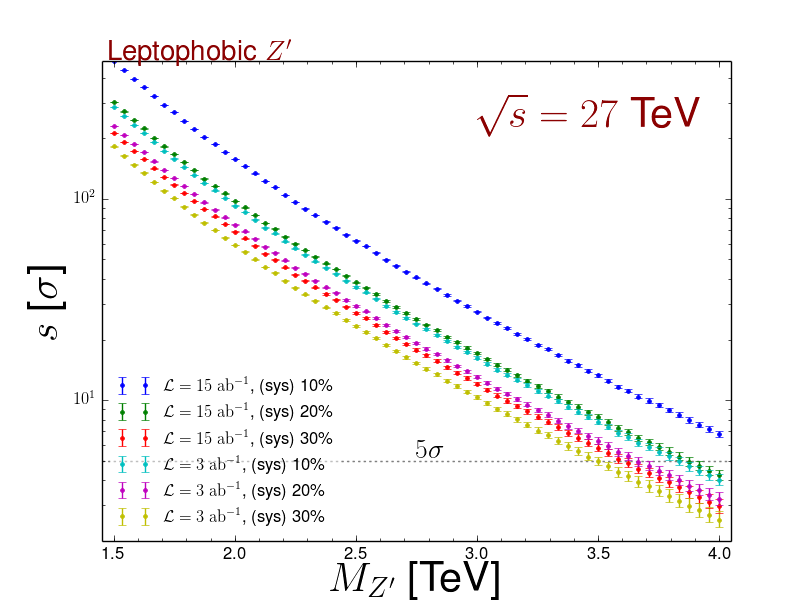}
 \includegraphics[scale=0.29]{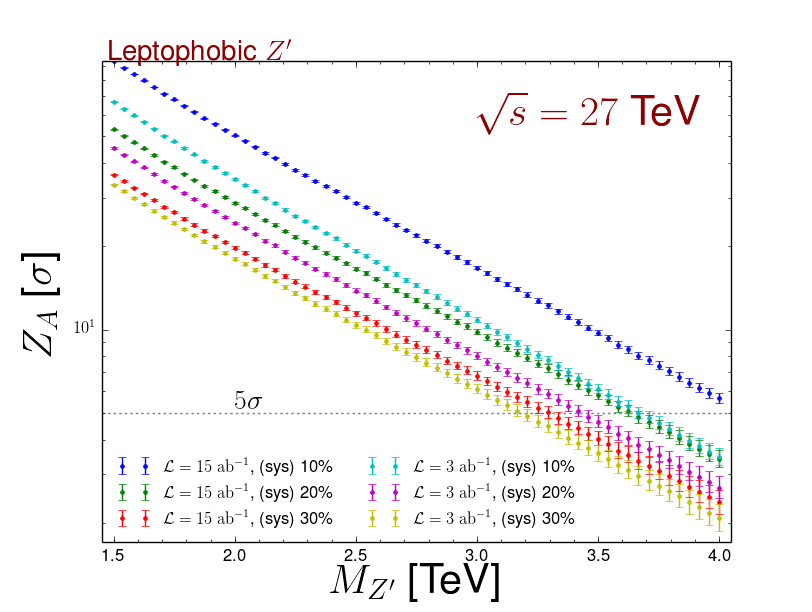}
 \caption{\it Left panel shows the standard significance against $ Z^\prime $ mass plotted for various luminosity and systematic uncertainty values. Right panel shows the same figure for Asimov significance, $ Z_A $. These results are presented in ref.~\cite{CidVidal:2018eel}. \label{fig:27tev}}
\end{figure}

It is possible to project these results into higher energies with different luminosity values to estimate the effect of HL-~\& HE-LHC. In \autoref{fig:27tev}, we calculated $ Z^\prime $ production cross section at $ \sqrt{s} = 27 $ TeV with its mass varying between $ 1.5- 4$ TeV. Using a naive assumption that all the kinematic structure and cut efficiencies will propagate in the same way, we projected our results to 3 ab$ ^{-1} $ and 15 ab$ ^{-1} $ and presented them with various systematic uncertainties on the background from 10\% to 30\% where the error bars showing the theoretical uncertainties. Our results shows that $ 5\sigma $ significance can be reached using standard ($ Z_A $) significance for a $Z^\prime$ at 3.5 (3.2) TeV assuming  30\% systematic uncertainty and 3 ab$ ^{-1} $ integrated luminosity. We showed that, due to the linear dependence of the systematic uncertainties to the luminosity, its effect increases with increasing luminosity.

\section{Conclusion}\label{sec:conclusion}

We investigate possible loopholes in $ E_6 $-inspired UMSSM searches at LHC, specifically through $ Z^\prime $ production. We showed that the inclusion of supersymmetric particles could cause a 20\% reduction in $ Z^\prime $ mass limits, and even those limits can be reduced further by changing the SUSY breaking scale. Throughout our analysis, we continually compared UMSSM with USM to emphasise the difference that inclusion of supersymmetric particles can create. 

Furthermore, we examined the possibility of having a leptophobic $ Z^\prime $ by adding kinetic mixing between $ Z-Z^\prime $ to avoid high-mass leptonic resonance searches.  We showed that the bounds from dijet resonances are considerably small and allows a light $ Z^\prime $ up to 1.6 TeV. To back up such claim, we proposed a supersymmetry driven channel to observe such leptophobic $ Z^\prime $. We showed that it is possible to reach $ 8\sigma $ sensitivity at 14 TeV centre-of-mass energy with 3 ab$ ^{-1} $ integrated luminosity using chargino production through $ Z^\prime $ where charginos further decayed into two leptons and missing energy. 

Assuming all cut efficiencies and kinematic structures will be preserved, we extrapolated our results to higher energies and luminosities. We showed the observable mass range for $ Z^\prime $ up to $ 5\sigma $ sensitivity for 27 TeV centre-of-mass energy. All results are compared with different systematic uncertainties on the background and different luminosity values to address the importance of understanding the background as well as the signal itself. These results show that investigating such scenarios are both feasible and worthwhile at HL-~\& HE-LHC.

%
%
\providecommand{\href}[2]{#2}\begingroup\raggedright\endgroup
\end{document}